\documentclass[reprint]{revtex4-1}

\usepackage{textcomp}
\usepackage{graphicx}
\usepackage[bitstream-charter]{mathdesign}
\usepackage{hyperref}
\usepackage{booktabs}

\newcommand{\mtext}[1]{\ensuremath{\textrm{\small#1}}}
\newcommand{\kev}[1]{\ensuremath{#1~\mathrm{keV}}}
\newcommand{\pA}[1]{\ensuremath{#1~\mathrm{pA}}}
\newcommand{\nA}[1]{\ensuremath{#1~\mathrm{nA}}}
\newcommand{\thresh}{\ensuremath{E_{\textrm{\scriptsize thresh}}}}
\newcommand{\vthresh}{\ensuremath{V_{\textrm{\scriptsize thresh}}}}
\newcommand{\qe}{\ensuremath{\mathit{QE}}}

\newcommand{\eph}{\ensuremath{E_\mtext{\scriptsize ph}}}
\newcommand{\muunit}[2]{#1~\textmu#2}

\begin{document}

\title{Characterization of an in-vacuum PILATUS~1M detector}
\author{Jan Wernecke}
\email[contact: ]{jan.wernecke@ptb.de}
\author{Christian Gollwitzer}
\author{Peter M\"uller}
\author{Michael Krumrey}
\affiliation{Physikalisch-Technische Bundesanstalt (PTB), Abbestr. 2-12, 10587 Berlin, Germany}

\begin{abstract}
    A dedicated in-vacuum X-ray detector based on the hybrid pixel PILATUS 1M detector has been
    installed at the four-crystal monochromator beamline of PTB at the electron storage ring
    BESSY~II in Berlin. Due to its windowless operation, the detector can be used in the entire
    photon energy range of the beamline from 10~keV down to 1.75~keV for small-angle X-ray
    scattering (SAXS) experiments and anomalous SAXS (ASAXS) at absorption edges of light elements.
    The radiometric and geometric properties of the detector like quantum efficiency, pixel pitch
    and module alignment have been determined with low uncertainties. The first grazing incidence SAXS
    (GISAXS) results demonstrate the superior resolution in momentum transfer achievable at low
    photon energies.
\end{abstract}

\maketitle

\section{Introduction}

    The advances of integrated circuits in the last few decades have significantly boosted the
    development of X-ray detectors~\cite{yaffe1997}. So-called hybrid pixel X-ray detectors have
    been developed, which consist of a readout chip bump bonded on a silicon sensor that acts as a
    radiation absorber~\cite{heijne1989}. State-of-the-art detectors like the
    XPAD~\cite{delpierre2007}, detectors based on the Medipix readout chip
    \cite{ponchut2001,pennicard2010}, and the PILATUS \cite{broennimann2006,kraft2009} combine a
    semiconductor pixel matrix with a readout chip providing an amplifier, comparator and digital
    counter for every single pixel. This is appealing especially for scattering and diffraction
    experiments, where the photon flux at individual pixels may vary over many orders of magnitude.
    As opposed to dose-proportional detectors, photon counting can provide very low darkcount rates,
    and consequently huge dynamic ranges, signal-to-noise ratios close to the quantum limit and
    negligible crosstalk between neighbouring pixels resulting in a nearly perfect point spread
    function~\cite{chmeissani2004}.

    The commercially available large-area (1 megapixel and above) hybrid pixel detectors are
    operated in air and the radiation enters through a thin window. This window limits the detectable
    photon energy range to energies above approximately \kev{5} due to absorption in the window. Yet
    the absorption edges of technologically and biologically relevant elements like silicon,
    phosphorus, sulphur, chlorine, or calcium are situated below this energy. To overcome this
    limitation, windowless operation in vacuum with a direct connection to the sample chamber is
    necessary. The suitability and performance of a PILATUS 100k detector under such conditions has
    been shown previously~\cite{marchal2011a,marchal2011b}. Moreover, extensive testing and
    characterization of detector modules in vacuum \cite{donath2013} has been carried out in
    collaboration with Dectris~Ltd.\ at the four-crystal monochromator (FCM) beamline in the
    laboratory of the Physikalisch-Technische Bundesanstalt (PTB) at the electron storage ring
    BESSY~II. However, these setups were preliminary experiments with a single module as a proof of
    concept at that time, a fully operational multi-module large-area in-vacuum PILATUS has not been
    realized up to now.

    In this paper, we fill this gap and describe the modifications made to the PILATUS~1M modular
    detector in collaboration with Dectris~Ltd.\ to operate under vacuum, so that the
    experimentally accessible energy range is widened downwards to a photon energy $\eph$ of
    \kev{1.75}. Radiometric as well as geometric characterization has been performed using traceable
    methods. The first measurement results in a typical scattering setup are reported to demonstrate the
    extended measurement capabilities at X-ray photon energies below \kev{4}.

\section{Experimental setup}

    \begin{figure}
        \caption{Motor axes of the SAXS setup. The axes for longitudinal detector movement
            $x_{\mathrm{det}}$ and for vertical movement of the stage $z_1$ and $z_2$ are equipped with
            optical encoders for absolute length measurements with micrometer resolution.}
            \includegraphics[width=0.8\columnwidth]{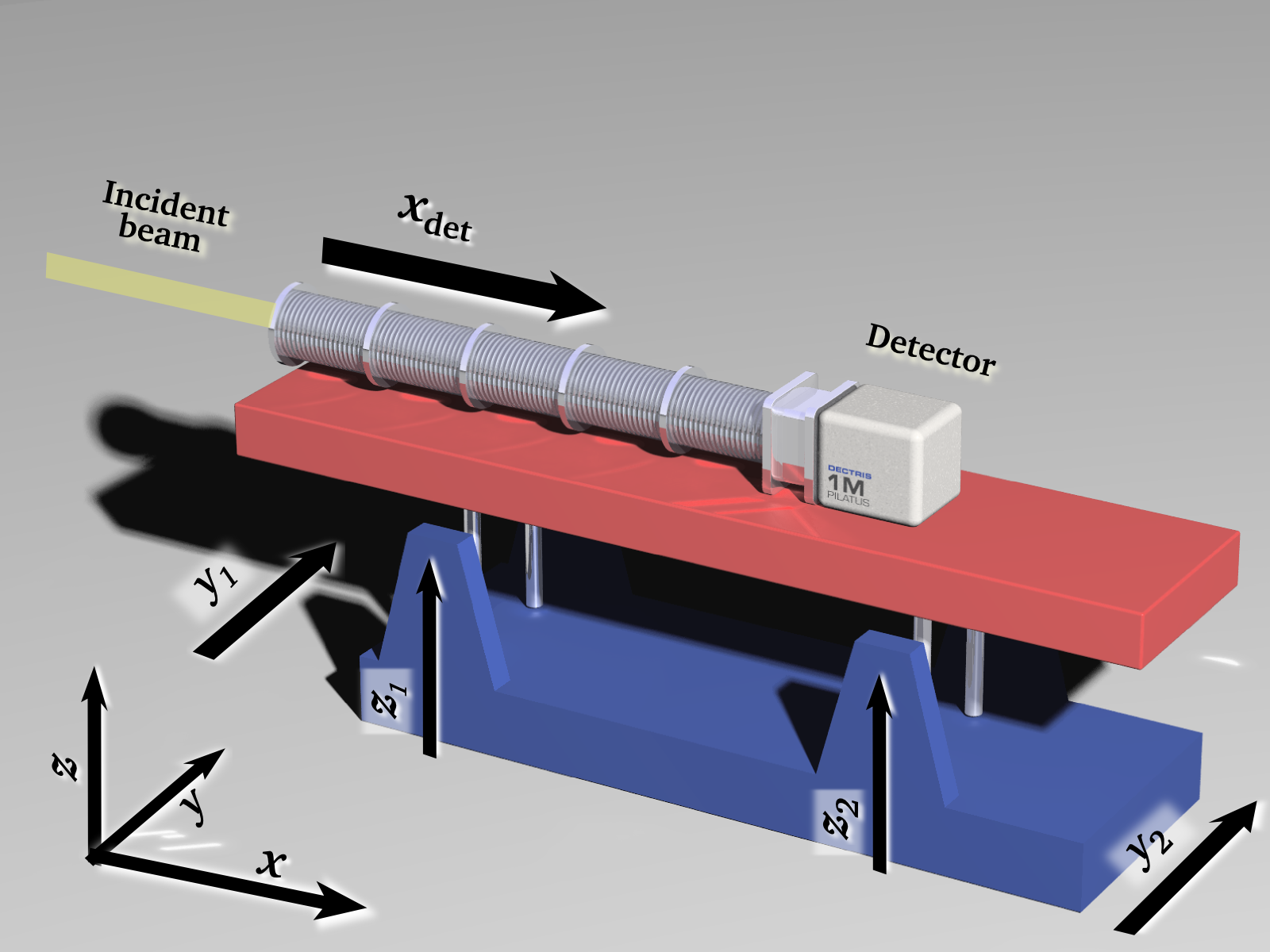}
        \label{fig:saxs}
    \end{figure}

    The in-vacuum PILATUS~1M detector was specifically developed and scaled to the parameters of the
    small-angle scattering setup of the FCM beamline \cite{krumrey2011,ptb-quartercentury}. The
    beamline \cite{krumrey2001} covers a photon energy range of \kev{1.75} to \kev{10}, which
    defines the targeted lower energy limit of the detector. The monochromator features an energy
    resolving power $\eph/\Delta \eph$ of $10^{4}$ and an accuracy of the energy scale of 0.5~eV
    \cite{krumrey2001,krumrey1998}. The photon flux of the incident beam at the location of the
    sample or detector can be measured in a traceable way by photodiodes that were calibrated
    against a cryogenic electric substitution radiometer \cite{gerlach2008} within a relative
    uncertainty of $1\ \%$.  A sample chamber equipped with six axes for sample movement is attached
    to the FCM beamline \cite{fuchs-newref}. For small-angle X-ray scattering measurements in
    transmission geometry (i.e., SAXS) and grazing-incidence reflection geometry (i.e., GISAXS), the 2D-detector
    is usually mounted on the SAXS instrument of the Helmholtz-Zentrum Berlin (HZB)
    \cite{hoell-saxs-patent} as illustrated in Figure~\ref{fig:saxs}. The detector is
    installed on a moveable stage (to the rear of Figure~\ref{fig:saxs}) and connected to an
    edge-welded bellow to allow any sample-to-detector distance between $1.75\ \mathrm{m}$ and
    about $4.5\ \mathrm{m}$, and a vertical tilt angle up to 3\textdegree\ without
    breaking the vacuum. The translation axes $z_1$, $z_2$, and $x_{\mathrm{det}}$ are equipped with
    optical encoders (Heidenhain AE~LC~182 and AE~LC~483) which measure the displacement on an
    absolute scale with an accuracy of \muunit{1}{m}.  These encoders establish the traceability of
    the detector displacement. The detector side of the bellow holds a moveable beamstop to block the
    intense transmitted or specularly reflected fraction of the beam.

\section{Technical implementation of the in-vacuum version}

    \begin{table}
        \caption{Technical specifications of the in-vacuum PILATUS~1M detector.}
        \label{tab:pilatus-tech-data}
        \begin{tabular}{ll}
            \toprule
            \multicolumn{1}{c}{parameter} & \multicolumn{1}{c}{value / setting} \\
            \midrule
            accessible photon energy            & \kev{1.75} \ldots $> \kev{36}$ \\
            sensitive detector area             & 179~mm $\times$ 169~mm \\
            sensor thickness                    & 320~\textmu m \\
            dimensions                          & ca. 60~cm $\times$ 37~cm $\times$ 37~cm \\
            mass                                & ca. 80~kg \\
            entrance flange                     & DN 250 CF \\
            typical cooler temperature          & 5\ \textdegree C \ldots 10\ \textdegree C \\
            typical operation pressure          & $< 1 \times 10^{-5}$~mbar\\ 
            pressure gauge                      & Pfeiffer PKR~251 \\
            \bottomrule
        \end{tabular}
    \end{table}

    \begin{figure*}
        \caption{(a) Sketch of the vacuum-compatible version of the PILATUS detector. The vacuum side
            (on the right) consists of the ten detector modules (grey) mounted on the downsized module carrier
            plate (blue). This is attached to the feed-through flange plate that contains the 575 electric
            connections and the water supply lines. The vacuum chamber (semi-transparent structure) is
            directly connected to the beamline. The air side (on the left) consists of the standard electronic
            units of the PILATUS and the water cooling supply. (b) Sketch of the feed-through flange plate
            that separates the vacuum and the air side and facilitates the electrical connection of detector
            modules.}
        \includegraphics[width=0.9\textwidth]{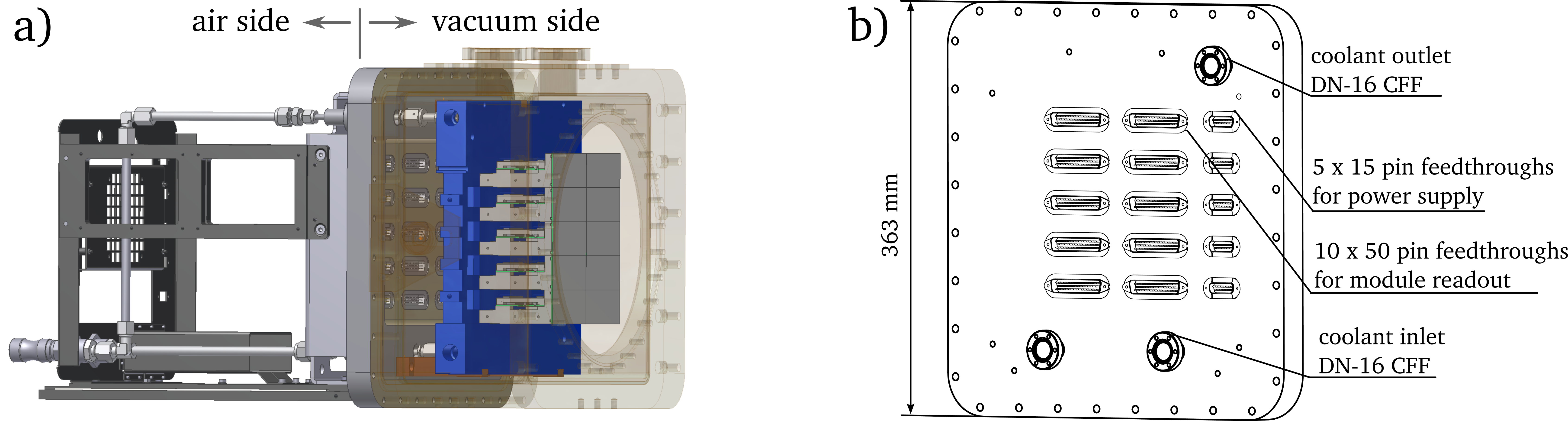}
        \label{fig:pilatus-setup}
    \end{figure*}

    One of the design goals for a vacuum-compatible version of the PILATUS~1M detector was to
    minimize the number of modifications from the standard detector. The final solution was a
    vacuum-proof separation of the detector modules from the electronic control units. To this
    end, a vacuum chamber for the modules and a feed-through flange plate were developed. The
    ten detector modules are mounted on a size-reduced module carrier plate.
    The carrier plate is connected to the feed-through flange plate which closes the vacuum chamber at the detector side.
    Figure~\ref{fig:pilatus-setup}a shows a sketch of the general setup.

    The \emph{vacuum chamber} encloses the detector modules that provide a total sensitive area of
    179~mm $\times$ 169~mm with a sensor thickness of 320~\textmu m. The CF-entrance flange has a diameter of 250~mm to prevent any shadowing
    of the detector surface and is directly connected to the HZB SAXS instrument.  A vacuum gauge is
    used for pressure monitoring and controls an interlock system, which shuts down the high voltage
    of the detector in case of vacuum loss.  The \emph{feed-through flange plate},
    Figure~\ref{fig:pilatus-setup}b, seals the vacuum chamber on the opposite side and facilitates
    the connection of the 575 electric lines and the channels for water cooling.  On the air side,
    standard PILATUS~1M electronic units are used for data processing.  The module carrier plate is
    cooled with circulating water kept at a constant temperature of typically 5~\textdegree C.
    Table~\ref{tab:pilatus-tech-data} gives an overview of the technical specifications of the
    in-vacuum PILATUS~1M detector.  Operation in air at higher photon energies is still possible
    with the modified PILATUS~1M setup. To this end, a Mylar window is attached to the
    entrance flange of the vacuum chamber.

    Before we describe the necessary electronic adjustments, the operation principle of the PILATUS
    hybrid pixel detector needs to be reviewed briefly. Many more details can be found in
    \cite{broennimann2006,kraft2009,kraft_characterization_2009}. The detection principle in each
    pixel is based on the generation of electron-hole pairs in a silicon pn-junction induced by an
    absorbed X-ray photon. The electric charge is amplified by a charge-sensitive preamplifier
    (CSA), the amplification of which can be set in discrete steps, which are called the \emph{gain
    modes} \cite{kraft_characterization_2009}. The amplified pulse is then compared to an adjustable
    threshold voltage \vthresh\ by a comparator. The pulse is registered and counted only if it
    exceeds the threshold, and otherwise discarded. The voltage threshold \vthresh\ corresponding to
    a photon energy threshold \thresh\ is determined by the software depending on the amplifier
    gain. In normal operation mode, the energy threshold \thresh\ is set to $\frac{1}{2}\  \eph$ to
    avoid charge-sharing counts in neighbouring pixels \cite{chmeissani2004}.

    For the in-vacuum PILATUS~1M detector, an additional ultra-high gain mode with higher
    amplification than the standard high-gain mode was added to account for the reduced number of
    electron-hole pairs generated by each photon at low X-ray photon energy. The lowest achievable
    \thresh\ is ultimately limited by amplifier noise exceeding the comparator threshold
    \vthresh\, or by the onset of instable operation. The minimum threshold determined is $\thresh =
    \kev{1.7}$ for stable operation in ultra-high gain mode. For the preferred threshold setting
    with $\thresh = \frac{1}{2}\  \eph$, this would only allow a minimal photon energy of
    \kev{3.4}.  In order to reach lower photon energies, for example, the silicon absorption K-edge
    at \kev{1.84}, \thresh\ can be set independently of the photon energy to a higher level. This
    results in a decreased count rate, but it also leads to a smaller effective pixel area because
    only photons that deposit at least a fraction of $\thresh / \eph$ of their energy in the pixel
    contribute to the counts \cite{schubert2010}. As a result, undersampling and aliasing occur
    which might even be an advantageous effect in some experiments, where a refined detector point
    spread function is needed~\cite{farsiu2004}.
    However, the usage of the ultra-high gain mode results in an increased detector dead-time of about 4~\textmu s, which results in a loss of registered photons \cite{marchal2011a}.

\section{Radiometric characterization}

    The quantum efficiency (\qe) of the detector, which is the ratio of registered counts to incident
    photons, was determined as a basis for measurements of absolute scattering intensities.  The
    \qe\ measurements were accomplished by taking sequences of images of the monochromatized
    synchrotron beam with varying energy. Before and after each sequence, the incident photon flux
    at each energy was determined by a calibrated photodiode.  The monochromatic photon flux of the
    beamline is in the order of $10^9\ \mathrm{s^{-1}}$ to $10^{10}\ \mathrm{s^{-1}}$ in an area of about
    $0.5\ \mathrm{mm^{2}}$ at the usual top-up ring current of $300\ \mathrm{mA}$ of the storage
    ring.
    This photon flux is well beyond the linear, unsaturated, detector response range, in particular in ultra-high gain mode and at low threshold energies.
    Hence, BESSY~II was operated in a special mode where the ring current was reduced stepwise to 832~\textmu A, 409~\textmu A, 95~\textmu A, and finally 6~\textmu A.
    This also allowed us to evaluate the linearity of the registered count rate in relation to the rate of incoming photons.
    The \qe\ was determined from the measurements at the lowest ring current, which resulted in photocurrents of the calibrated diodes from \pA{14} to \nA{1.2} (darkcurrent \pA{<1}).
    Additionally, the beam was defocused so that the most intense spot covered an area of approximately 100 pixels. 
    In this way, the maximum flux of incoming photons and pixel was kept below 20\,000~$\mathrm{s}^{-1}$, while the minimum photon flux in the evaluated region of 10~$\mathrm{s}^{-1}$ still exceeded the darkcount rate of $10^{-5}\ \mathrm{s}^{-1}$ by several orders of magnitude.

    \begin{figure}
        \caption{Registered counts per second along the most intense region of the illuminated area (see inset for a logarithmic image of the spot shape, the white line indicates the cut line of the plotted profiles). The detector images where recorded at four different storage ring currents under otherwise identical conditions (\eph = \kev{2.5}).
        Each profile has been scaled by the ratio of minimal ring current (6~\textmu A) and the ring current of the profile.
        Deviations from a linear counting behaviour manifest in lower scaled count rate in comparison to the 6~\textmu A profile.}
        \includegraphics[width=0.99\columnwidth]{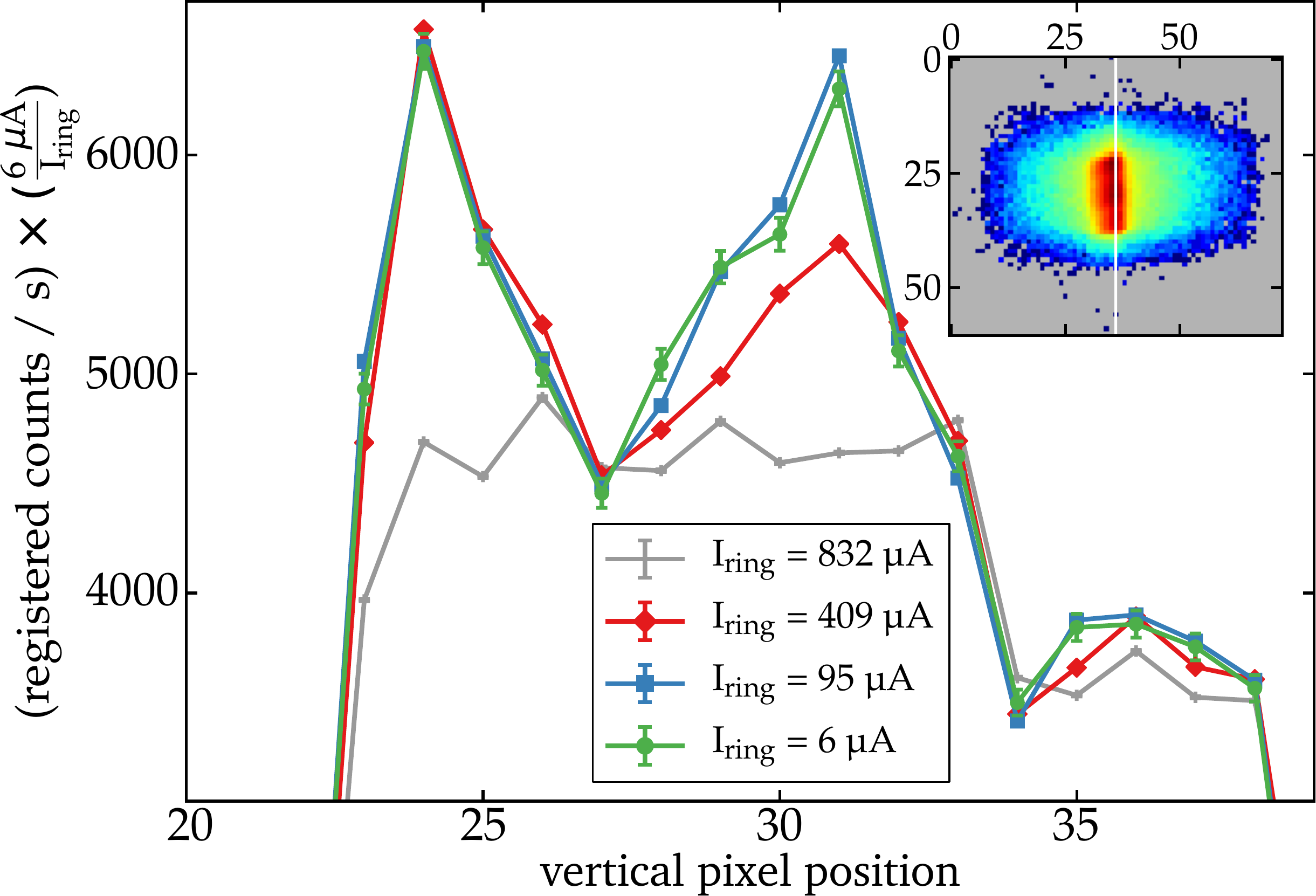}
        \label{fig:linearity}
    \end{figure}

    Before the \qe\ can be accurately determined, the linear response of the detector must be checked and the uncertainty contributions need to be evaluated.
    Displayed in Figure~\ref{fig:linearity} are the registered counts per second and per pixel along the most intense line of the illuminated area (see vertical line in the inset) for the four different ring currents, each recorded under otherwise identical conditions at $\eph = \kev{2.5}$.
    The profiles have been scaled by the ratio of the ring current of the \qe\ measurements (6~\textmu A) to the corresponding ring current of the profile.
    In this way, an increase of detector saturation due to a too high rate of incoming photons (which is equivalent to the ring current) can be observed by a deviation from the unscaled count rate profile measured at 6~\textmu A.
    It can be seen that the profiles of 832~\textmu A and 409~\textmu A deviate significantly from the 6~\textmu A profile, clearly indicating the occurrence of saturation.
    But the profile of 95~\textmu A differs by less than 2.2~\% from the 6~\textmu A data, which should give an upper estimate for the increase of saturation from 6~\textmu A to 95~\textmu A.
    The \qe\ measurements were carried out at a ring current of 6~\textmu A, where even a much lower deviation from the linear counting behaviour can be expected.
    Nonetheless, we use a relative uncertainty contribution of 2~\% to the \qe\ measurement as an upper estimate for the effect of nonlinear counting.
    The contribution of the uncertainty of the photon energy of $u(\eph)/\eph = 10^{-4}$ is negligible.
    The comparison of photo diode measurements before and after each set of PILATUS measurements yields a mean deviation of 0.5~\%.
    In conjunction with the uncertainty of the diode calibration, this yields a relative uncertainty of 1~\% of the incoming photons flux.
    In total, the resulting relative uncertainty of the \qe\ in ultra-high gain mode, in particular at low photon energies below \kev{4}, is 3~\%.
    In high gain mode, the incoming photon flux is well within the linear regime.
    Therefore, the corresponding relative uncertainty in this setting is only determined by the variation of before-and-after measurements with the photodiodes, which is within 1~\%.
    
    \begin{figure*}
        \caption{The quantum efficiency \qe\ of the detector, measured over the full energy range of the
            beamline with recommended settings (a) and for different threshold levels at the low end of
            the energy range (b).  The green triangular symbols in (a) denote measurements using the
            high-gain mode, while all other data was measured using the ultra-high gain mode.
            The shaded areas around the data points indicate the relative uncertainty of the values (3~\% in ultra-high gain mode, 1~\% in high gain mode).
            The inset in (b) displays a close-up around the silicon K-edge.
            The ring current of the storage ring was reduced to \muunit{22}{A} (\muunit{6}{A}) for photon energies above (below) $3.5\ \mathrm{keV}$, respectively.}
        \includegraphics[width=0.9\textwidth]{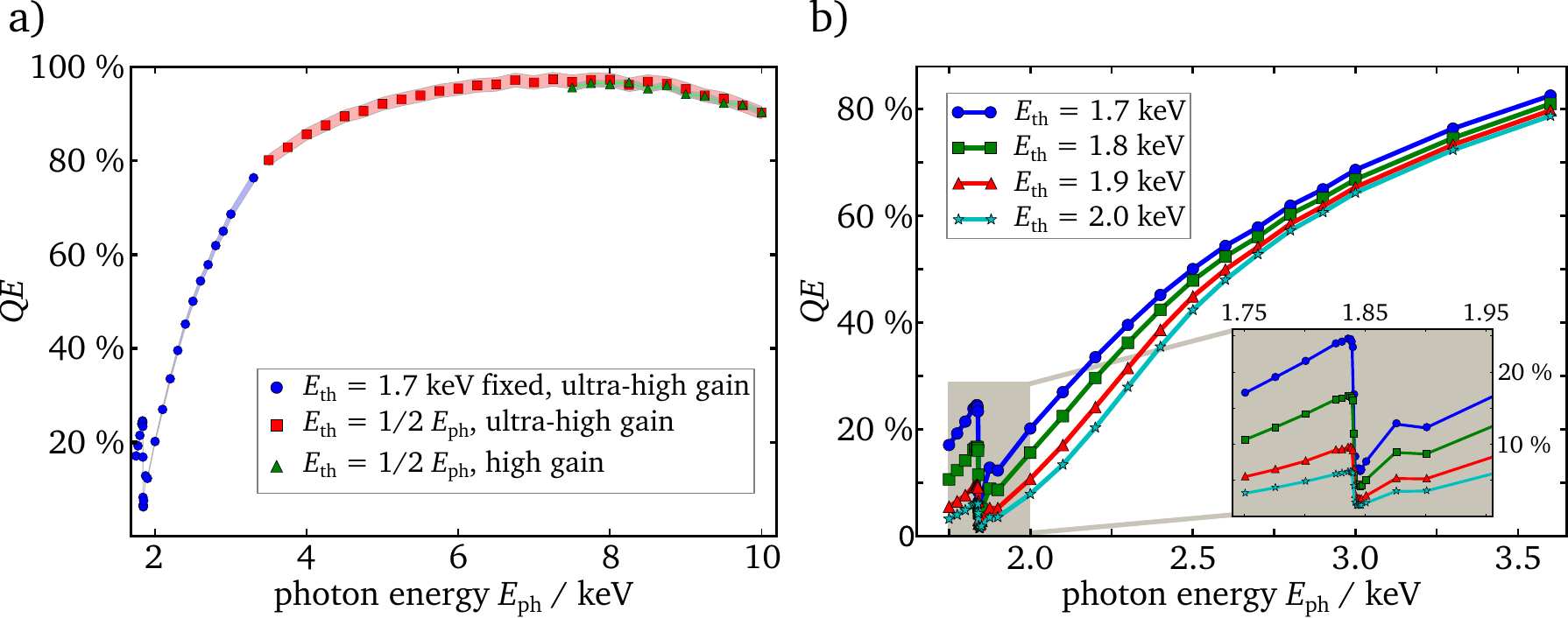}
        \label{fig:qe}
    \end{figure*}

    The measured quantum efficiency with the associated uncertainty (shaded areas) is displayed in Figure~\ref{fig:qe}. It depends not only on the
    photon energy \eph, but also on the threshold level \thresh\ of the detector.
    Above $\eph=\kev{3.4}$, the threshold level was set to the preferred value
    $\thresh=\frac{1}{2}E_\mtext{ph}$, which is shown in Figure~\ref{fig:qe}a by the red square symbols for
    the ultra-high gain mode and by the green triangles for the high-gain mode. The high-gain mode is
    limited to threshold settings \thresh\ above \kev{3.75}, or equivalently \eph\ to above \kev{7.5}.
    Below $\eph=\kev{3.4}$, the threshold in ultra-high gain mode was fixed to $\thresh=1.7\
    \mathrm{keV}$ (blue circles in Figure~\ref{fig:qe}a). In addition, the \qe\ was measured in this range
    for larger settings of \thresh\ up to \kev{2.0} (Figure~\ref{fig:qe}b).

    The \qe\ exceeds $80\ \%$ over the range from \kev{3.4} to \kev{10}, with a maximum of $96\ \%$
    at $8\ \mathrm{keV}$. Below \kev{3}, the quantum efficiency is reduced due to the absorption of
    photons
    in the non-sensitive surface layers of the sensor, which are always present in semiconductor
    detectors \cite{krumrey1992self}. Just above the Si K-edge, the \qe\ drops to about $5\ \%$, however,
    measurements are feasible down to \kev{1.75}.
    The measured \qe, in particular at low energy, is in full agreement with the previously reported \qe\ of the single module test setup at the corresponding threshold setting \cite{donath2013}.
    The two different gain settings result in an
    difference less than $1\ \%$, which is within the uncertainty of the measurement. The threshold level
    settings have a noticeable influence, as displayed in Figure~\ref{fig:qe}b. The highest \qe\ is
    achieved by the lowest possible threshold setting $\thresh = \kev{1.7}$, as
    expected~\cite{kraft2009}, and is therefore chosen as the recommended setting for all subsequent
    measurements.

    It should be noted that the fill pattern of the electrons in the storage ring also has an influence on the registered count rate as described by \cite{trueb2012}.
    During our measurements, the circulation period was 800~ns, the electrons were divided in 350 bunches with a separation time of 2~ns and a dark gap of 100~ns.
    Since the detector dead-time in ultra-high gain mode of 4~\textmu s corresponds to more than four cycles, the detector is completely insensitive to the fill pattern substructure.
    The fill pattern during our measurements is comparable to the data obtained at the Swiss Light Source and at the Australian Synchrotron as reported by Trueb et al., therefore, a similar systematic loss in the registered count rate should occur.

    \begin{figure*}
        \caption{The homogeneity of the detector at a photon energy of (a) \kev{10} and (b) \kev{5} in ultra-high gain mode and with $\thresh = \frac{1}{2}\, \eph$.
            The raw data was preprocessed by binning $10\times 10$ pixels into one in order to overcome
            the quantum noise in comparison to the inhomogeneous detector response. The ring in the
            centre is an artefact which comes from the positioning of the beamstop to the centre of the 
            scattering pattern.}
        \includegraphics[width=0.80\textwidth]{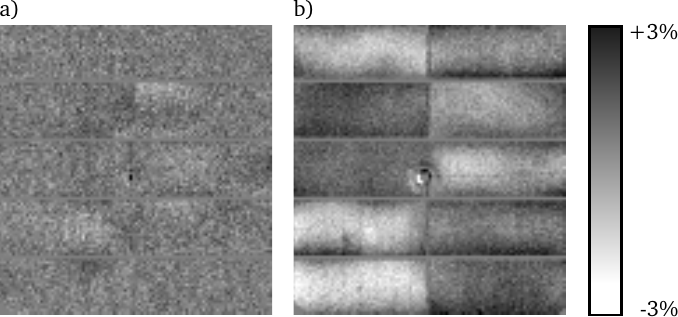}
        \label{fig:homogeneity}
    \end{figure*}

    In order to investigate possible variations in sensitivity over the detector area for fixed
    settings of energy, gain mode and threshold, a different approach was applied. SAXS images in
    the range from
    \kev{4} to \kev{10} were recorded using a sample of glassy carbon. The scattering pattern of glassy
    carbon exhibits a flat plateau in the range of the momentum transfer $q$ from
    $0.1\ \mathrm{nm^{-1}}$ to $1\ \mathrm{nm^{-1}}$~\cite{zhang2010}.  Therefore, we achieve a nearly
    homogeneous illumination of the detector, which varies only in radial direction from the scattering
    centre. By dividing the whole image by the azimuthally averaged scattering curve pixel by pixel, we
    obtain an image with the relative deviation of each pixel value from the mean.
    Figure~\ref{fig:homogeneity} displays the intensity deviation after averaging patches of
    $10\times 10$ pixels in order to reduce the shot noise.
    At \kev{10}, the intensity difference amounts to $0.5\ \%$ across the whole detector, while
    at \kev{5}, the intensity varies by $2.5\ \%$, although the manufacturer-supplied flat field
    correction was enabled. This discrepancy can be explained by the absorption of radiation in the
    upper insensitive layer of the detector. At high energies, this layer is nearly transparent,
    while at lower energies, the absorption and therefore the variation increase.
    This may result in a limited accuracy of the extrapolation of calibration values for trimming, which is based on flat field reference measurements at higher photon energies.
    The inhomogeneity can possibly be reduced by applying better flat field corrections in the low photon energy range from these images.

\section{Geometric characterization}

    A possible geometric distortion introduced by the detector must be known to determine uncertainty
    bounds for metrological nanodimensional measurements such as suggested, for example, in
    \cite{krumrey2011,wernecke2012}.  A sequence of measurements was conducted in order to determine the
    pixel pitch, the displacement of the modules from their nominal position and the misalignment with
    respect to one another. This was achieved by measurement sequences, where the small-angle scattering
    of a selected sample was used to generate static test patterns, and the detector was moved to
    different positions for each image of the sequence.

    The first sequence of measurements was conducted in SAXS geometry at $\eph = \kev{8}$ using the
    standard sample silver behenate, which displays an intense ring at
    $q=1.076\ \mathrm{nm^{-1}}$ \cite{blanton1995}. The detector was positioned at a distance of
    $d=2754\ \mathrm{mm}$ to the sample, and $240$ images were recorded. Between the exposures, the
    detector was vertically shifted in a stepwise fashion by moving both vertical translation axes $z_1$
    and $z_2$ in parallel.  The total distance by which the detector was moved amounts to
    $7\ \mathrm{mm}$. The traceability of the $z_1$ and $z_2$ movement was established by the Heidenhain
    linear encoders.

    \begin{figure*}
        \caption{Test patterns for geometric measurements (a) SAXS image of silver 
            behenate at 8~keV together with fit circle in dashed yellow.
            (b) GISAXS pattern (similar to Fig.~\ref{fig:gisaxs}) to determine the module alignment.
            Image series were recorded for vertical and horizontal displacement of the detector, here shown exemplary for a vertical movement along a module gap (grey shaded area).
            The numbered maxima indicate the peaks that were tracked to determine the detector movement (arrows show the nominal path along which 20 images were recorded).
            One group of peaks (6-12; green arrows) crosses the module borders, the other group (1-3 and 15-17; red arrows) stays on a module and is used for reference.
        The same procedure is applied to the other module gaps on the detector.}
        \includegraphics[width=0.80\textwidth]{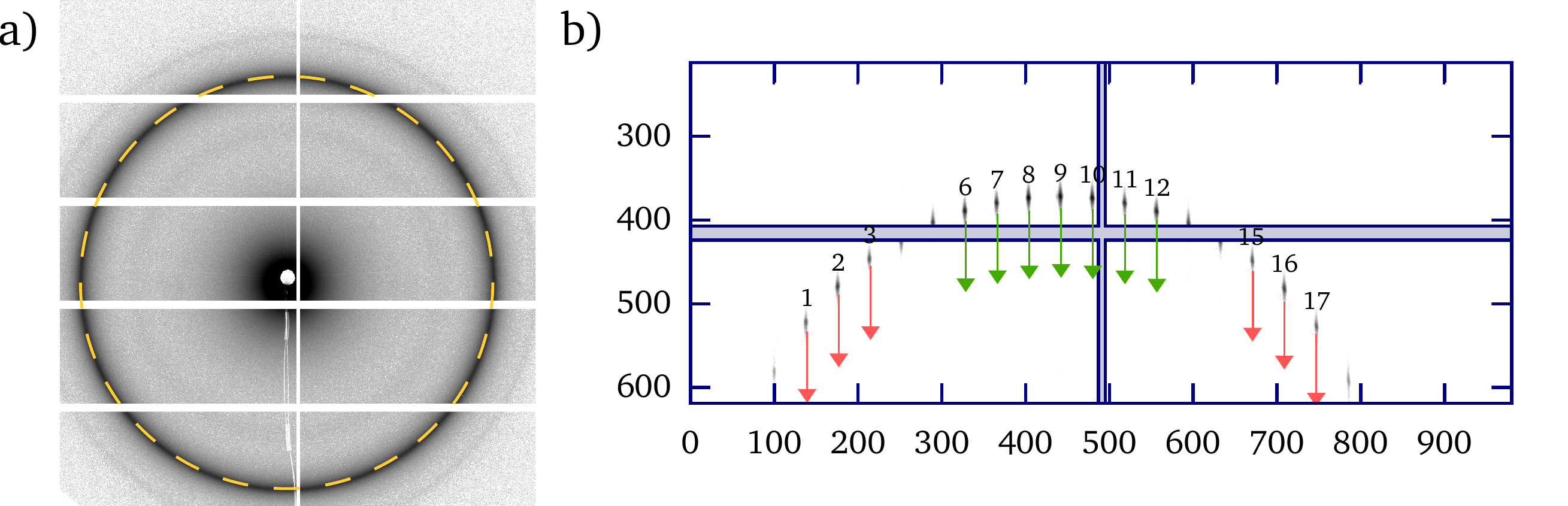}
        \label{fig:geometrymeasurement}
    \end{figure*}

    Next, a circle was fitted to every recorded image by maximizing the average intensity along the
    ring, which was represented by a Gaussian line with a width of $\sigma=1.06\ \mathrm{px}$.  An
    example image together with the fitted circle (dashed line) is shown in
    Figure~\ref{fig:geometrymeasurement}a. The best fit centre positions of these circles were then
    linearly fitted to the corresponding vertical detector displacement  values $z_1$ and $z_2$.  The
    residuals for this fit did not exceed one tenth of the pixel pitch for any circle position. From this
    linear fit, the pixel pitch $p=(172.1\pm 0.2)$\muunit{}{m} can be concluded.  The uncertainty
    estimate of this value is derived from the comparison of both vertical shift axes and two
    independent measurements. The pure statistical error from the linear fit is smaller by an order of
    magnitude.

    For the second sequence of measurements, the GISAXS pattern of a reflection grating with
    parallel alignment of grating lines and incident beam was used (see next section and
    \cite{wernecke2012}). This setup produces a series of equidistantly spaced sharp peaks ordered
    on an extended semicircle, which was used to characterize the placement of the individual
    detector modules with respect to each other. Figure~\ref{fig:geometrymeasurement}b displays the
    positions of the peaks on the detector for one contiguous series of images. The detector was
    moved vertically upwards in 20 steps and an image was taken at every position. The peak
    positions were extracted from the images with subpixel resolution by computing the
    intensity-weighted centre of mass for every peak. The peaks can be divided into three
    categories. The peaks in the first category (labelled 1--3 and 15--17 in
    Figure~\ref{fig:geometrymeasurement}b) stay on a single module. These were used as a reference
    trace. The peaks 6--10 and 11--12 cross the horizontal module borders between the upper and
    lower modules to the left and right, respectively.  The remaining peaks (unlabelled) are neither
    confined to a single module nor do they cross the module borders completely to reach the next
    module. Similar datasets were recorded for all module borders in the horizontal and vertical
    direction.

    The relative displacement of the modules from the nominal position results in a discontinuity of
    the trace for the border-crossing peaks.  However, on the subpixel scale it has to be considered
    that the movement of the detector is slightly irregular due to deviations of the mechanical
    positioning.  By comparing the border-crossing traces with the reference peak traces, the
    discontinuity can be detected regardless of an irregularly shaped path. The analysis was
    performed by least-squares fitting of the reference trace to the border-crossing traces at both
    sides of the gap. The maximum deviation from the nominal position amounts to \muunit{60}{m} over
    the whole detector, which is less than 1~pixel.

    In principle, the same method could be used to determine the in-plane angular misalignment
    between two neighbouring modules. The angular deviation was found to be below 0.1\textdegree,
    but this is already beyond the limit of this method due to the limited resolution of the peak-
    centre finding of $\approx 20$\muunit{}{m}. An out-of-plane angular misalignment leads only to
    smaller pixel length in the direction perpendicular to the axis of rotation. A deviation was
    measured for the same detector with great sensitivity by Bragg diffraction at the surface of the
    detector~\cite{gollwitzer2013}, with a result of a deviation of at most 0.4\textdegree. However,
    we cannot distinguish whether the deviation originates from a possible miscut of the silicon wafers or
    from a mechanical misalignment of the modules. Since the cosine of this angle deviates by less
    than 25~ppm from unity, this has no effect on the scattering images.

\section{Application example: GISAXS at low photon energies}

    One of the advantages of a lower X-ray photon energy in SAXS and GISAXS experiments is the
    increased resolution (at a given experimental geometry).  Consequently, the scattering pattern
    of larger structures can be resolved and the precision in determining smaller scattering lengths
    increases due to a larger separation distance of scattering features. In X-ray scattering, the
    reciprocal space is mapped. In SAXS and GISAXS, this is manifested in an intensity pattern of the
    diffusely scattered beam that is recorded by the 2-dimensional detector.  For GISAXS, the
    relevant momentum transfer coordinates are
    \begin{equation}
        q_y =  k_0 \; \left( \sin\theta_f \cos\alpha_f \right) ;\ 
        q_z =  k_0 \; \left( \sin\alpha_f  + \sin\alpha_i \right)
        \label{eq:q}
    \end{equation}
    with the wavenumber $k_0 =  2\pi / \lambda$, the incident angle $\alpha_i$, and the vertical and
    horizontal scattering angles $\alpha_{f}$ and $\theta_f$, respectively. From (\ref{eq:q}) it
    becomes clear that a reduction of photon energy (i.e., increase of wavelength $\lambda$) at a
    given geometry results in a decrease of the probed $q$-range and an increase of the
    $q$-resolution of the detector image.  This has high practical relevance in nanometrological
    GISAXS measurements of sub-\textmu m and nm-spaced gratings \cite{wernecke2012,hofmann}. The aim
    of such measurements is to establish a traceable determination of the grating period, line
    width, and other structural parameters.  This may serve as a basis to evaluate the general
    accuracy of the GISAXS method itself and gives more meaning to any length determined with GISAXS
    by an associated uncertainty. Here, the benefit of an increased $q$-resolution is a reduction of
    the grating parameter uncertainty.

    \begin{figure*}
        \caption{GISAXS scattering pattern of a line grating with a period length of 833~nm in parallel
            orientation of grating lines and incident beam, recorded at (a) \kev{8} and (b) \kev{3}
            (incident angle 0.8\textdegree\ in both cases). (c) Intensity profiles along the
            semicircles of the GISAXS patterns. (d) Close-up of the $q_{y}$-range from
            $-0.11~\mtext{nm}^{-1}$ to $-0.01~\mtext{nm}^{-1}$.}
        \includegraphics[width=0.98\textwidth]{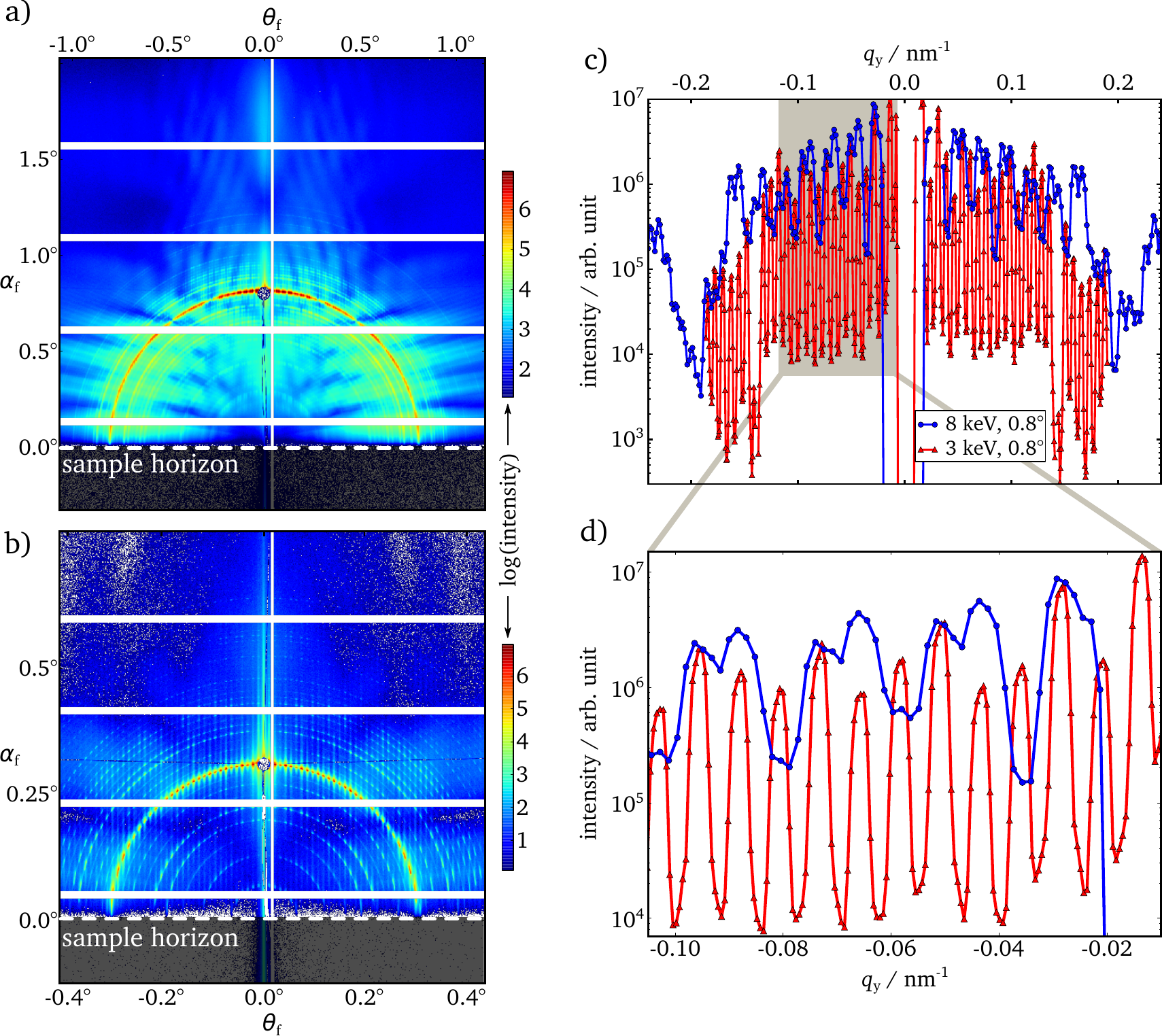}
        \label{fig:gisaxs}
    \end{figure*}

    Figure~\ref{fig:gisaxs}a shows a typical GISAXS pattern for parallel orientation between the
    incident X-ray beam and the grating lines with an incident angle of $\alpha_i$ = 0.8\textdegree.
    The most prominent feature is a semicircle with evenly spaced intensity maxima. The modulation
    of these maxima is governed by the characteristic scattering lengths that are present in the
    sample \cite{mikulik2001,yan2007}. Hence, by analysing the frequencies of the intensity profile
    along the semicircle, the period length and line width of the grating can be directly
    determined from the scattering pattern \cite{wernecke2012}. The GISAXS images recorded at 8~keV,
    Figure~\ref{fig:gisaxs}a, and at 3~keV, Figure~\ref{fig:gisaxs}b, already show the enlarged
    separation distance of maxima at the lower energy. Figure~\ref{fig:gisaxs}c shows the
    intensity profiles along the semicircle as a function of $q_y$ for both energies. The profiles
    and the close-up of the region left of the beamstop (at around $q_y$ = 0~nm$^{-1}$) in
    Figure~\ref{fig:gisaxs}d show the significantly pronounced oscillations and the increased number
    of data points per peak at 3~keV. This allows a more precise identification of the
    oscillation frequencies of the signal, which in turn results in lower uncertainties of the
    structural grating parameters determined.

\section{Conclusion}

    A vacuum-compatible version of the PILATUS 1M detector has been installed at the PTB
    four-crystal monochromator beamline and enables scattering measurements down to a photon energy
    of $1.75\ \mathrm{keV}$, which is below the K-absorption edge of silicon and other light
    elements also relevant for biological and organic systems. The quantum efficiency has been
    determined in the entire range provided by the FCM beamline with a relative uncertainty of $3\
    \%$ in ultra-high gain mode and $1 \%$ in high gain mode. The quantum efficiency is excellent ($>80\ \%$) above $3.4\ \mathrm{keV}$ and provides a
    sufficient signal for X-ray scattering measurements at lower photon energies down to $1.75\
    \mathrm{keV}$. The geometric distortions of the detector due to deviations in module placement
    stay below 1 pixel over the whole detector. The first scattering experiments show the extended
    capabilities of the detector due to the increased resolution in $q$ at low energies. SAXS and
    GISAXS measurements on biological samples and nanostructured polymer thin films are currently
    being analysed and will be published soon. Further insight into the internal structure is expected from
    the element-selective tuning of the scattering contrast of the contained light elements.

\acknowledgments
    We want to thank Levent Cibik and Stefanie Marggraf (PTB) for their extensive support during the
    installation and characterization of the detector. We would also like to acknowledge Tilman
    Donath, Pascal Hofer, and Benjamin Luethi (Dectris~Ltd.) for helpful discussions and advice
    during the setup and characterization. The technical assistance of the HZB machine group of
    BESSY~II, who set the ring current to the reduced levels for the radiometric measurements is
    appreciated, as well as the cooperating research with the HZB SAXS instrument with Dr. Armin
    Hoell.

\bibliographystyle{aipnum4-1}
\bibliography{refs}

\end{document}